\documentclass[onecolumn]{aastex631}

\begin{document}

\def\msol{\hbox{$\hbox{M}_\odot$}}
\def\lsol{\hbox{$\hbox{L}_\odot$}}
\def\kms{km s$^{-1}$}

\title{Statistical Properties of the Population of the Galactic Center Filaments:\\
The Spectral Index and Equipartition Magnetic Field
}

\author[0000-0001-8403-8548]{F. Yusef-Zadeh} 
\affiliation{Dept Physics and Astronomy, CIERA, Northwestern University, 2145 Sheridan Road, Evanston , IL 60207, USA
(zadeh@northwestern.edu)}

\author[0000-0001-8403-8548]{R. G. Arendt} 
\affiliation{Code 665, NASA/GSFC, 8800 Greenbelt Road, Greenbelt, MD 20771, USA}
\affiliation{UMBC/CRESST 2 (Richard.G.Arendt@nasa.gov)}

\author[0000-0001-8403-8548]{M. Wardle} 
\affiliation{Dept of Physics and Astronomy,  Research Centre for Astronomy, Astrophysics
and Astrophotonics, Macquarie University, Sydney NSW 2109, Australia, (mark.wardle@mq.edu.au)}

\author[0000-0001-8403-8548]{I. Heywood} 
\affiliation{Astrophysics, Department of Physics, University of Oxford, Keble Road, Oxford, OX1 3RH, UK}
\affiliation{Department of Physics and Electronics, Rhodes University, PO Box 94, Makhanda, 6140, South Africa}
\affiliation{Department of Physics and Electronics, Rhodes University, PO Box 94, Makhanda, 6140, South Africa 
(ian.heywood@physics.ox.ac.uk)}

\author[0000-0001-8403-8548]{W. Cotton} 
\affiliation{National Radio Astronomy Observatory, Charlottesville, VA, USA, (bcotton@nrao.edu)}

\author[0000-0001-8403-8548]{F. Camilo} 
\affiliation{South African Radio Astronomical Observatory, 2 Fir Street, Black River Park, Observatory,
Cape Town, 7925, South Africa, (fernando@ska.ac.za)}

\begin{abstract} 
We present high-pass filtered continuum images of the inner $3.5^\circ\times2.5^\circ$ of the Galactic center at 20 cm with $6.4''$ resolution. These
mosaic images are taken with MeerKAT and reveal a large number of narrow filaments, roughly an order of magnitude increase in their numbers compared
to past measurements. For the first time, we carry out population studies of the spectral index and magnetic field of the entire region. 
The mean spectral
indices of the filaments are steeper than supernova remnants (SNRs) (-0.62) with a value of $\alpha\sim-0.83$.
The variation
in $\alpha$ is much larger than for the SNRs, suggesting that these characteristics have a different origin. A large-scale cosmic-ray driven wind has
recently been proposed to explain the origin of filaments and the large-scale 430 pc bipolar radio and X-ray structure.  This favors the possibility
that the large-scale bipolar radio/X-ray structure is produced by past activity of Sgr A* rather than coordinated burst of supernovae. A trend of
steeper indices is also noted with increasing distance from the Galactic plane. This could be explained either by synchrotron cooling or weak shocks
accelerating cosmic-ray particles in the context of the cosmic-ray driven wind. The mean magnetic field strengths along the filaments ranges from
$\sim100$ to 400 $\mu$G depending on the assumed ratio of cosmic-ray protons to electrons. 
Given that there is a high cosmic ray pressure in the Galactic center, the large equipartition magnetic field implies that the magnetic filed is
weak in most of the interstellar volume of the Galactic center.
\end{abstract}


\section{Introduction} 
\label{sec:intro}

More than 35 years have passed since the discovery of  the magnetized radio filaments associated with the Galactic center 
Radio Arc near $l\sim0.2^\circ$ \citep{zadeh84}.  These observations showed linear, magnetized features running 
mainly perpendicular to the Galactic plane. Their morphology was unique and different than shell-like or 
jet-like nonthermal radio continuum sources that had been observed.  This was the first indication that the 
nucleus of our Galaxy harbors energetic activity that produces  relativistic particles along straight filaments with no 
obvious source of acceleration. Dozens of nonthermal radio filaments  with similar characteristics to the prototype 
filaments in the Radio Arc  have been discovered in the intervening years indicating narrow synchrotron structures and are 
confirmed by polarization measurements
\citep{liszt85,zadeh86,gray91,haynes92,staguhn98,lang99a,lang99b,larosa01,larosa04,zadeh04,nord04,law08,pound18,staguhn19,arendt19}.


Past radio continuum surveys have generally focused on the morphology and characteristics of individual filaments. 
Spectral index  
and polarization characteristics of limited number of filaments have been studied 
\citep{inoue84,zadeh86,zadeh87,gray95,tsuboi95,zadeh97,lang99a,lang99b,wang02,law08,pare19}. The spectral indices of a dozen 
filaments indicate  a wide range of values between +0.2 and -1.3 
\citep{law08}.  Given the  evidence for an abundance of clusters of 
filaments throughout 
the Galactic center (Heywood et al. 2019), we carried out a study that statistically characterizes 
the  properties of the 
population of filaments  based on data presented in Heywood et al. (2022, in press). 
Unlike previous studies that have focused on individual filaments  with the exception of survey works by \cite{law08,nord04}, here we determine 
the distribution of the spectral index  and the equipartition  magnetic field   of the entire population of 
radio filaments using 1.28 GHz in-band analysis of filtered data.   
A more detailed account of geometrical properties of the filaments such as position, length, position angle, and curvature, as well as the median 
brightness of each filament are given elsewhere (Yusef-Zadeh et al. 2022, in preparation).

 
\section{Data Reduction}
\label{sec:data}

Details of the MeerKAT observations can be found in Heywood et al. (2022, in press). Here we briefly describe the 
observations and the data processing. The mosaic is constructed from 20 MeerKAT pointings of $\sim$7.2 hours each 
(excluding calibration scans) for a total of 144 hours on-source. The L-band (856$-$1712~MHz) system was used, with 
the correlator configured to deliver 4,096 frequency channels, which were averaged by a factor of 4 prior to 
processing. The 1.28 GHz mosaic covers the inner $\sim$3.5$^{\circ}$~$\times$~$\sim$2.5$^{\circ}$ ($l$~$\times$~$b$) 
of the Galactic center, with some aggressive tapering of the data delivering an angular resolution of 4$''$, with a 
map pixel extent of 1.1$''$. The data were also imaged in 16 sub-bands, with Gaussian tapering applied to the 
visibilities in order to try to match the angular resolution across the band. An inner cut to remove spacings 
shorter than 164 wavelengths was also applied when imaging the sub-bands, in order to prevent the lower frequencies 
detecting large angular-scale structures that are invisible to the interferometers at higher frequencies. The 320 
individual images (16 sub-bands times 20 pointings) were visually examined, with images that were significantly 
affected by radio frequency interference being discarded, resulting in 269 images in total. Prior to production of a 
mosaicked cube, we convolved the model images with a circular 8$''$ Gaussian restoring beam, and summed the result 
with the residual images convolved with an appropriate homogenization kernel. A spectral index mosaic was then 
produced by fitting for the slope of the log-frequency, log-brightness spectrum for every sight line through the 
mosaicked cube. To ensure we are producing robust spectral index values we enforce the criterion that more than 8 
frequency bands must be present for a fit to be performed, as well as masking the resulting spectral index mosaic in 
regions where the total intensity brightness was less than 1~mJy beam$^{-1}$. Lastly, 
VLA images at 20cm based on two pointings,  centered on Sgr A* and Sgr C G359.5+0.0,  were used to make astrometric corrections, as discussed in detail 
in  Heywood et al. (2022).



\subsection{High-pass Filtering}
	

One of the challenging issues in high dynamic range radio imaging of the Galactic center is the non-uniform 
background on a wide range of angular sizes across the  large-scale mosaic image. 
 This creates difficulties in 
characterizing the properties of linear filaments in a standardized manner, in particular measuring the 
spectral index along the length of long filaments. 
The angular scales and amplitude variations of the background make it important
to subtract it before measuring brightnesses and spectral indices of the
filaments. However, the general complexity and size of the field make it
impractical to select and subtract backgrounds in a means that involves user
defined selection of background. 
To enhance the visibility of the  filaments, the mosaic 
image is filtered using a difference of Gaussians to smooth noise and remove large scale backgrounds. The original images are 
convolved with large and small Gaussian beams before they were subtracted from each other.  
This is 
essentially a spatial frequency filtering of the image, which is commonly used as an edge-enhancement technique. 
The process was implemented as a single convolution using the kernel shown in Figure 1. The 
narrower, smoothing Gaussian function (red curve in Fig. 1) 
has $\sigma_1=2.5$ pixels with pixel size of 1.1$''$. The wider, 
background-subtracting Gaussian function (inverted blue curve in Fig. 1) has $\sigma_2=4.5$ pixels. These values were subjectively determined 
as ones that helped highlight the filamentary features in the image as well as their spectral index 
distribution.  Point sources in the final filtered image have FWHM $\sim 6.4''$. 
Together the two Gaussians form the black kernel in Figure 1, which was convolved with the original image. 
The units of the resulting map are still reported in Jy 
beam$^{-1}$, where the beam area is that of the original beam. The process reduces the numerical value of the brightness in 
filament  pixels by a factor of $\sim$0.13, with large variations due to the effect of the background removal. 
This factor is not applied to the filtered image intensities and does not effect the calculation of spectral indices, 
but it is applied in estimating magnetic field strengths (Section 3.2).
 This filtering is applied not only to the mosaic image but also to the cube of 16 frequency 
channels before  the spectral index image is constructed following the prescription that is described in the previous section.

\section{Results} 
\label{sec:filter}

For general orientation, Figure 2a shows the original mosaic image at a resolution of 4$''\times4''$ with prominent 
features (Sgr A to Sgr E) labeled. A number of previously studied individual filaments, HII regions, SNRs,  and other well-known 
radio sources  such as the Mouse toward this 
region have also been labeled (see \cite{yb87,larosa01,zadeh04,nord04,law08} and references therein).  Figure 2b shows the 
corresponding filtered 
image where spatially varying diffuse background emission has been subtracted, leaving a background 
with a standard deviation of $\sim3$ $\mu$Jy beam$^{-1}$ across the roughly 60\% source-free portions of the field.
For this reason, numerous fainter filaments are uncovered and stand out in the filtered image. The identification of 
faint features  is more difficult in the unfiltered image because of the contamination by diffuse and varying background 
emission. However, to confirm the reality of faint filaments in the filtered image, we compared and found their counterparts
 in the unfiltered 
image. There are a large number of twisted short ``spaghetti-like" features noted along the Galactic plane. These are 
associated mainly with HII regions and SNRs and have not been used in the study of filaments having different 
morphology, lengths,  and distributions away from the Galactic plane.  


The filaments appear to be  pervasive throughout this region. 
Given the abundance of new radio filaments in a relatively small volume, it is difficult to manually localize the 
individual filaments. 
Results presented here are independent of the identification and properties of individual filaments. 
However,  we have used filtered images to automatically construct 
a catalog of individual filaments with their characteristics which  will be published elsewhere. 


\subsection{Spectral Index Distribution}

Figure 3a shows the spectral index image of the entire filtered image whereas Figure 3b  selects  filaments that are 
longer than 132$''$. These images show   
 three distinct classes of objects with different spectral indices. 
There are HII regions showing flat spectra between --0.2 to --0.3 (yellow to red)
such as the Sgr B complex, the Arches,  and diffuse extended
emission along the Galactic plane. SNRs such as the Sgr A East
SNR G0.0+0.0, SNR G359.1-0.5, SNR G359.0-0.9, SNR 0.33+0.04, SNR G0.9+0.1 and Sgr D SNR
\citep[e.g.,][]{green88,kassim96}, have a typical spectral index
$\alpha\sim-0.6$ to $-0.7$ (green to blue). Lastly, there are nonthermal radio
filaments that have broad spectral index distribution as well as  
the steepest spectral index that  can readily  be discerned  at
high latitudes (green to blue) with typical $\alpha\sim-0.8$.  
There are exceptions such as the flat spectral index of the Radio Arc with $\alpha$ ranging between $-0.2$ to $0.2$. The flat spectrum of the Radio Arc
is consistent with numerous past measurements \citep{anantha91,sofue92,pohl92,reich00,wang02,nord04,law08,ludo16,pound18,staguhn19}.


Figure 4 presents the histograms of the spectral index of filaments  longer than 66$''$ and 132$''$, respectively. 
In Figure 4a, there is a strong narrow peak at $\alpha \approx -0.14$, which 
represents the remaining thermal emission features. The  much broader peak with $\alpha \approx -0.62$ represents nonthermal synchrotron emission from SNRs (with some contribution from the 
filaments). The selection criterion of $L> 132''$ excludes essentially all thermal features and many of the filaments associated with SNRs. 
This histogram in Figure 4b is dominated by the nonthermal filaments and indicates a mean spectral index of $\alpha \approx -0.83\pm0.44$ which is
generally steeper than the typical spectral index of SNRs (above). The fitted mean spectral index values are consistent with trends expected from SNRs,
HII regions and nonthermal filaments. The standard deviations of the spectral indices of these different components reflect intrinsic dispersion in the
emission from these sources, rather than measurement errors.

The  lengths ($L$) of the filaments that are selected  in Figures 3 and 4  are measured by 
application of the procedure $LOOPTRACING\_{AUTO}$ \citep{aschwanden10}; (Yusef-Zadeh et al. 2022).  
Length criteria are applied  to reject  shorter structures of HII regions and SNRs throughout the 
Galactic plane. The $L>66''$ selection criterion excludes features that are essentially background. 

(Yusef-Zadeh et al. 2022).  Length criteria are chosen to increasingly reject 



Another notable trend is a steepening of the spectral index as a function of
absolute Galactic latitude. Figures 5a and 5b show this trend for lengths of filaments longer than 66$''$ and $132''$, respectively.
In the former plot, a large number of thermal sources with flat spectral indices obscure the trend, but selecting
only the longest filaments (i.e., mostly nonthermal filaments) the trend is clearer despite a large intrinsic scatter.




\subsection{Equipartition Magnetic Field}

The magnetic field for each pixel is determined by using the filtered synchrotron intensity.  Assuming a spectral index of $\alpha=-0.5$ produced by an $E^{-2}$ electron spectrum running from 10\,MeV to 10\,GeV,  and neglecting the presence of relativistic protons and nuclei, the estimated equipartition magnetic field scales as
\begin{equation}
B_{\rm eq} (\mathrm{mG})  = 0.172  \,(9.21\,j_\nu(\mathrm{mJy/arcsec}^3)  )^{2/7}
\end{equation}
where the synchrotron emissivity $j_\nu$ is related to intensity $I_\nu$ by $j_\nu = I_\nu/ L$ and $L$ is the depth of the source.  Here we have adopted a Galactic center distance of 8\,kpc in converting between length and angular scale (i.e. $\approx0.04$\,pc/arcsec).  The filament width is typically unresolved in the filtered image and is assumed to be $4''$;  the depth of a typical filament along the line of sight is also taken to be the same as its width,  $4''$.

We also calculated the equipartition magnetic field accounting for the observed spectral index.  We first computed the synchrotron emissivity $j_\nu = I_\nu/L$ at 1.18 and 1.38\,GHz for a grid of assumed field strengths B$_{rm eq}$  and  electron power-law indices $p$ (where $n(E)\propto E^{-p}$ between 10\,MeV and 100\,GeV).   We averaged the fluxes to obtain an effective 1.28\,GHz flux density and used their ratio to compute the spectral index.  We then fit the results to obtain an empirical expression for the equipartition field given the observed flux and spectral index:
\begin{equation}
    B_{rm eq}(\mathrm{mG}) =  a\, ( j_\nu / j_0 )^b 
\end{equation}
where
$a = (0.5145 + 0.9756\alpha + 1.0377\alpha^2 ) / ( 1 - 1.1192\alpha - 0.2749\alpha^2),\, 
b = 1 / (3.0584 - 0.8597\alpha)$, 
and $j_0$ = 125.4 $\mu$Jy/arcsec$^3$.  

As expected, the above two estimates of the equipartition magnetic field almost coincide when $\alpha=-0.5$.  Figure 6 shows  $B_{rm eq}$ as a function of spectral index for different adopted ranges of relativistic electron energies. The estimated field strength is sensitive to the adopted lower cutoff for steep spectra, and the upper cutoff for flat spectra.   This behavior is driven by the electron energy density implied by the assumed spectra.    A related uncertainty is the ratio of the proton to electron energy densities: the estimates above assume that this is zero but $B_{rm eq}$ increases  by almost a factor of 4 if p/e$=100$.

Figure 7a shows the equipartition magnetic field strength across the entire field, assuming $\alpha=-0.5$ (Eq. (1)). Figure 7b is similar to 7a 
except that the more detailed estimate of Eq. (2) is applied, and thermal sources are masked by choosing spectral indices that are steeper than 
$\alpha=-0.4$. This is effective at removing background thermal emission and H II regions (e.g. Sgr B), but also does exclude some flat spectrum 
filaments, most notably in the Radio Arc at $l\sim 0.2\arcdeg$.  Figure 7b 
shows that many identified filaments have estimated fields of the order of 100 $\mu$G (blue color) and a maximum value of $\sim400\mu$G.
Figure 8a and 8b show the histograms of the equipartition magnetic field values for 
long and  the longest filaments 
 estimated from the filtered and masked image of Figure 7b. The peak for $L>66''$ and $L>132''$
filaments is $\sim0.2$ mG.

\section{Discussion} 
\label{sec:filter}

Our statistical study described in the previous section estimated typical values that characterize 
 the spectral index  and the equipartition magnetic field distributions of 
the population of Galactic center radio filaments.  The in-band spectral index values  are measured   simultaneously over 800 MHz bandwidth  
centered at 1.28 GHz. 
The equipartition magnetic  field values are derived  by incorporating the dependence of 
$B_{\rm eq}$ on the spectral index. 
These derivations and measurements  focus on  global mean values over a large region of the Galactic center. 

Previous studies have used widely separated radio bands and focused on individual spectral indices and equipartition magnetic fields 
of individual  filaments
\citep{lang99a,lang99b,sofue92,liszt95,pohl92,wang02,larosa04,nord04,law08,gray91,gray95}.  The largest number of spectral indices were given by 
\citep{law08} using 6 and 20 cm wavelengths. Our spectral indices  of individual filaments generally agree with previous 
measurements in spite of different spectral windows, sensitivities, and techniques of background subtraction. 
For example, the spectral index of the Snake filament G359.1-0.2 as a function of latitude is estimated in Figure 3b and \citep{gray95}. 
We note that the northern half of the G359.1-0.15, north of $b=-0.18'$, has a  flatter spectrum than the rest of the 
filament. The 
spectral index at $b=-10'$ agrees with that estimated by \citep{gray95}, though their spectral index gradient values 
 do  not agree with our work 
 because of  their limited 20cm beam size, as pointed out by  \citep{gray95}. 
In another example,  the steepening of the spectral index away from the Radio Arc 
 is  consistent  with  one group of filaments noted by  \citep{pohl92}. 
Previous estimates of the equipartition magnetic field as a function of the spectral index 
 give values that 
are consistent with our derivations. The magnetic field of the Snake filament is estimated to be 88 $\mu$G  \citep{gray95}
which is similar to that shown in Figure 7b.


\subsection{The Spectral Index of the Filaments and SNRs: The Origin of the Bipolar  Radio Bubble}


We determined the mean spectral index of the population of filaments indicating a steeper 
spectrum on average ($\alpha = -0.83\pm0.44$) than SNRs ($= -0.62\pm0.5$).  The quoted uncertainties are the standard deviations of the histogram 
components in Figure 4 and do not include systematic errors such as those introduced by the automated global procedures applied here, which introduce 
a degree of contamination in the samples (see Yusef-Zadeh et al. 2022).  The comparison between these two populations of 
nonthermal sources in the Galactic center is relevant because of their potential role in explaining the origin of large-scale bipolar radio and X-ray 
structures. The triggering event that has been proposed to explain the origin of the radio bubble and the hot X-ray emitting coronal gas is either 
enhanced accretion onto the 4$\times10^6$ \msol\, black hole, Sgr A*, or a burst of star formation activity a few times $10^5-10^6$ years ago 
\citep[e.g.][]{heywood19,ponti19}.  However, the fact that the number of filaments declines with latitude and that the filament population is mainly 
bounded by the radio structure, suggests a causal connection between the filaments and the radio bubble \citep{heywood19}.
 If so, scenarios for the radio bubble must also explain the origin of the filaments.  The distinction between the spectral indices of SNRs 
and filaments disfavors synchronized SNe and implies an outburst from Sgr A* is a more likely explanation of the origin of the radio bubble, the filaments, and the X-ray chimney \citep{ponti19,heywood19}.

Another property of the spectral index distribution is its trend of steepening with increasing absolute Galactic latitude. There are two possible 
explanations.  On one hand, the trend could result from synchrotron cooling, which steepens the synchrotron spectrum as the underlying electron 
population ages.  In this picture the relativistic populations of electrons illuminating the filaments at extreme latitudes are older than those 
closer to the Galactic plane.  If the filaments arise as a consequence of interacting cosmic-ray driven winds emerging from the Galactic plane, the 
filaments at higher latitudes are produced when cosmic-ray particles in the winds have aged more than those particles closer to the Galactic plane.  
The energy loss rate depends on the magnetic field along the filaments. If we assume the magnetic field is 30 or 100 $\mu$G, the synchrotron cooling 
time scale is 5.5 or 0.9 Myr, respectively, at the observed frequency $\nu=1.2$ GHz.  These time scales are consistent with the inferred age of the 
outburst that created the radio bubble and X-ray Chimney.  On the other hand, if shocks driven by a large scale cosmic-ray driven wind as it flows 
around obstacles (such as stellar wind bubbles) are responsible for the production of individual filaments, then the shocks produced at higher 
latitudes must be weaker, thus resulting a steeper energy spectrum of particles.

\subsection{The Magnetic Field Strengths of the  Filaments and  the Galactic Center ISM}

We estimated the mean magnetic field of the population of filaments to be in the range $\sim$100-400 $\mu$G, possibly larger, 
 depending on the proton to electron 
ratio. This is significantly larger than estimates of the weak magnetic field permeating the central 100\,pc of the Galaxy, which range from $\sim$10\,$ 
\mu$G to ~$30\,\mu$G \citep{larosa05,zadeh13}. The amplification of the magnetic field occurs along string-like filaments, implying that most of the 
interstellar volume of the Galactic center has to have lower magnetic field than that of the filaments, implying lower magnetic energy densities than 
the observed high cosmic ray pressure in this region, as discussed in the next section.  We assumed equipartition between particles and the magnetic field along the 
filaments.  The 
high cosmic ray pressure observed throughout the Galactic center 
suggests that it is unlikely that the diffuse  magnetic field is  in equipartition with the field along the filaments.
Diffuse nonthermal structure in the Galactic center 
imposes other limitations on the 
large-scale magnetic field strength, suggesting that the pervasive magnetic field is also  weak, on the order of 10$\mu$G \citep{larosa05} and 
not of the order of $\sim$1 mG as suggested by \cite{morris94,chandran00}. 
Additional support comes 
from earlier measurements of the rotation measure distribution of extragalactic radio sources projected toward the Galactic center \citep{roy04},  
and with observations of the Snake filament \citep{gray95}, which indicates a weak  magnetic field strength in  the Faraday 
screen toward the Galactic center. 
Lastly, the weak diffuse field  is also consistent with a turbulent model of the Galactic center
in which the  magnetic field originates from turbulent activity. 
In this picture, the spatial distribution of the magnetic field energy is highly intermittent, 
and the regions of strong field have filamentary structure whereas the large-scale diffuse magnetic field is weak \citep{boldyrev06}.

\vfill

\subsection{The Origin of Excess Cosmic-ray Particles in the Galactic Center Region}

One of the key questions that remains is the origin of the elevated cosmic ray ionization rate in the Galactic center ISM. The evidence for this 
comes from the H$_3^+$ absorption line, the heating rate of warm molecular gas, ionization by nonthermal emission, and 6.4 keV line studies.  The 
mid-infrared H$_3^+$ absorption line measurements imply that the central 100\, pc of the Galaxy is permeated by low-density molecular medium with 
density $n_\mathrm{H} \sim100$ cm$^{-3}$ subject to a cosmic ray ionization rate $\zeta_\mathrm{H}\sim 10^{-14}$\,s$^{-1}$\,H$^{-1}$, a value that is 
consistent with other indicators, such as the temperature of Galactic center molecular clouds. Estimates of the cosmic ray ionization rate from these 
studies are two to three orders of magnitudes greater than elsewhere in the Galaxy 
\citep[e.g.][]{geballe99,oka05,indriolo12,lepetit16,zadeh07a,zadeh07b,zadeh13,goto14,oka19}.

The high cosmic ray pressure was recently suggested to drive a large-scale wind away from the Galactic plane creating the bipolar X-ray  and radio 
emission. In this picture, the non-thermal radio filaments result from the interaction of the large-scale wind and stellar wind bubbles 
\citep{zadeh19}.  The correlation of nonthermal radio filaments and the radio bubble was first noted in Heywood et al. (2019). The nonthermal 
filaments are magnetized streamers created by the wrapping of the wind's magnetic field around an obstacle such as a source of stellar wind blowing 
in the ISM, by analogy with the magnetized ion tails of comets embedded in the solar wind.  Recent MHD simulations of the propagation of cosmic rays 
along harp-like systems of filaments is consistent with the suggestion that cosmic ray streaming is preferred over diffusion along the filaments 
\citep{thomas20}.

What is the mechanism that injects excess cosmic ray particles and produces diffuse nonthermal emission from the inner few hundred parsecs of the 
Galactic center? One possibility is that the enhanced global ionization rate is a relic of past outburst activity, either of correlated supernovae 
associated with a burst of star formation, or of the central supermassive black hole Sgr A*.  These scenarios can explain the origin of the radio 
bubble, the hot X-ray emitting gas and the filaments \citep{heywood19,zadeh19}.

Alternatively, the enhanced cosmic-ray population in the Galactic center region might be supplied by leakage or escape from the radio filaments.  In this 
scenario, the power in energetic particles accelerated and subsequently leaked by the filaments into their surrounds must be sufficient to compensate 
for the energy losses associated with ionization throughout the Galactic center, as well as the losses by outward advection in a Galactic nuclear 
wind.

To explore this we first estimate the power required to maintain the cosmic-ray population at the Galactic center.  The inferred ionization rate 
implies a cosmic ray energy density $\sim1000$\,eV\,cm$^{-3}$ -- a thousand times that at the solar neighborhood.  This population is subject to 
ionization losses of $\chi \sim 10$\,eV per ionization, yielding a net loss rate per unit volume $\sim \zeta_\mathrm{H}n_\mathrm{H}\chi \approx 
10^{-11}$\,eV\,cm$^{-3}$\,s$^{-1}$.  In turn, this implies a lifetime against ionization losses $\sim 0.3$\,Myr.  This is comparable to estimates of 
the loss time by advection in a Galactic center wind that has been invoked to explain the X-ray and radio bubbles extending above and below the Galactic 
plane. This would advect particles out of the plane at speeds of order 400\,km\,s$^{-1}$, yielding a loss time on 100\,pc scales of order 0.25\,Myr.  
In total, the loss rate is $\approx 2\times 10^{-11}$\,eV\,cm$^{-3}$\,s$^{-1}$.

Now we determine whether the filaments can supply this power.  First, the relativistic particle energy density in the filaments is of order the 
magnetic field energy density, which for our estimated mean field in the filaments $B\sim 0.3\,$mG, is $B^2/8\pi\approx 2000$\,eV\,cm$^{-3}$.  For a 
particle leakage time scale $\tau$ and a volume filling factor $f$, the volume-averaged energy loss rate from the filaments is $fB^2/(8\pi\tau) 
\approx 2\times 10^{-11} fB_{0.1}^2/\tau_\mathrm{Myr}$\,eV\,cm$^{-3}$, implying a requirement that the leakage time scale $\tau\approx f$\,Myr, i.e. 
a few times $10^4$\,yr for $f$ of a few percent.  

To explore if  the filaments are able to supply the cosmic rays responsible for the high 
ionization rate found in the diffuse ISM of the Galactic center, Figure 9a and 9b show  histograms of 
 the number of filaments as a function of median surface
brightness for linear features longer than $>66''$ and $> 132''$, respectively. The red lines indicate 
power law fits to the bright portion of the histograms, indicating  power law
indices of  of $k=-0.75$ and $k=-0.68$, respectively. Using these power law distributions of the brightness of the filaments to extrapolate to 
much fainter filaments,  
the effective filling factor is estimated to be  less than a percent, thus we expect short leakage time scale. This implies that   
the filament population may not be powerful enough to flood the Galactic center  region with cosmic rays (Yusef-Zadeh et al. 2022).

\subsection{Conclusion}

In summary, we present an analysis of the entire system of non-thermal radio filaments in the Galactic center and 
determine the mean values of important physical quantities in this region. In particular, we use filtered images to 
statistically determine the spectral index and the equipartition magnetic field of the population of nonthermal 
radio filaments. The steep spectral indices of the filaments compared to SNRs and their steepening away from the 
Galactic plane support a picture of an outburst from Sgr A* than correlated supernovae associated with a burst of 
star formation. We also estimated that the mean equipartition value of the magnetic field of the population of 
filaments is strong. An argument was made that the amplification of the magnetic field occurs along the filaments, 
which implies that most of the interstellar volume of the Galactic center where the CMZ resides must have weak 
magnetic fields. We also suggested that one mechanism for the leakage of energetic particles from filaments may not 
able to explain the high cosmic ray ionization rate noted in the Galactic center region. In this picture, the 
leakage results from the interaction of filaments with low density diffuse gas throughout the Galactic center. 
Alternately, the interaction of nonthermal filaments with denser clouds has been argued in the past to explain: the 
origin of the warm temperature of molecular clouds, and the 6.4 keV line emission from neutral FeI from dense 
clouds, as well as diffuse $\gamma$-ray emission from diffuse gas \citep{zadeh13}. In this picture, diffusion of 
cosmic ray particles will be different for low and high energy cosmic ray particles. While low energy cosmic-ray 
electrons loss is significant, high energy particles can penetrate deep into dense molecular clouds without any 
losses.


\begin{acknowledgments}
We thank the referee for useful comments. Work by R.G.A. was supported by NASA under award number 80GSFC21M0002. FYZ is partially supported by the grant AST-0807400 from the the 
National Science Foundation. The MeerKAT 
telescope is operated by the South African Radio Astronomy Observatory, which is a facility of the National Research Foundation, an agency of the Department of Science and 
Innovation. The authors acknowledge the Center for High Performance Computing (CHPC), South Africa, for providing computational resources to this research project. 
The National Radio Astronomy Observatory is a facility of the 
National Science Foundation operated under cooperative agreement by Associated Universities, Inc. IH acknowledges support from the UK Science and Technology Facilities Council 
[ST/N000919/1], and from the South African Radio Astronomy Observatory which is a facility of the National Research Foundation (NRF), an agency of the Department of Science and
Innovation
\end{acknowledgments}

\facilities{VLA, MeerKAT}

\bibliographystyle{aasjournal}


\begin{figure}[ht!]
\plotone{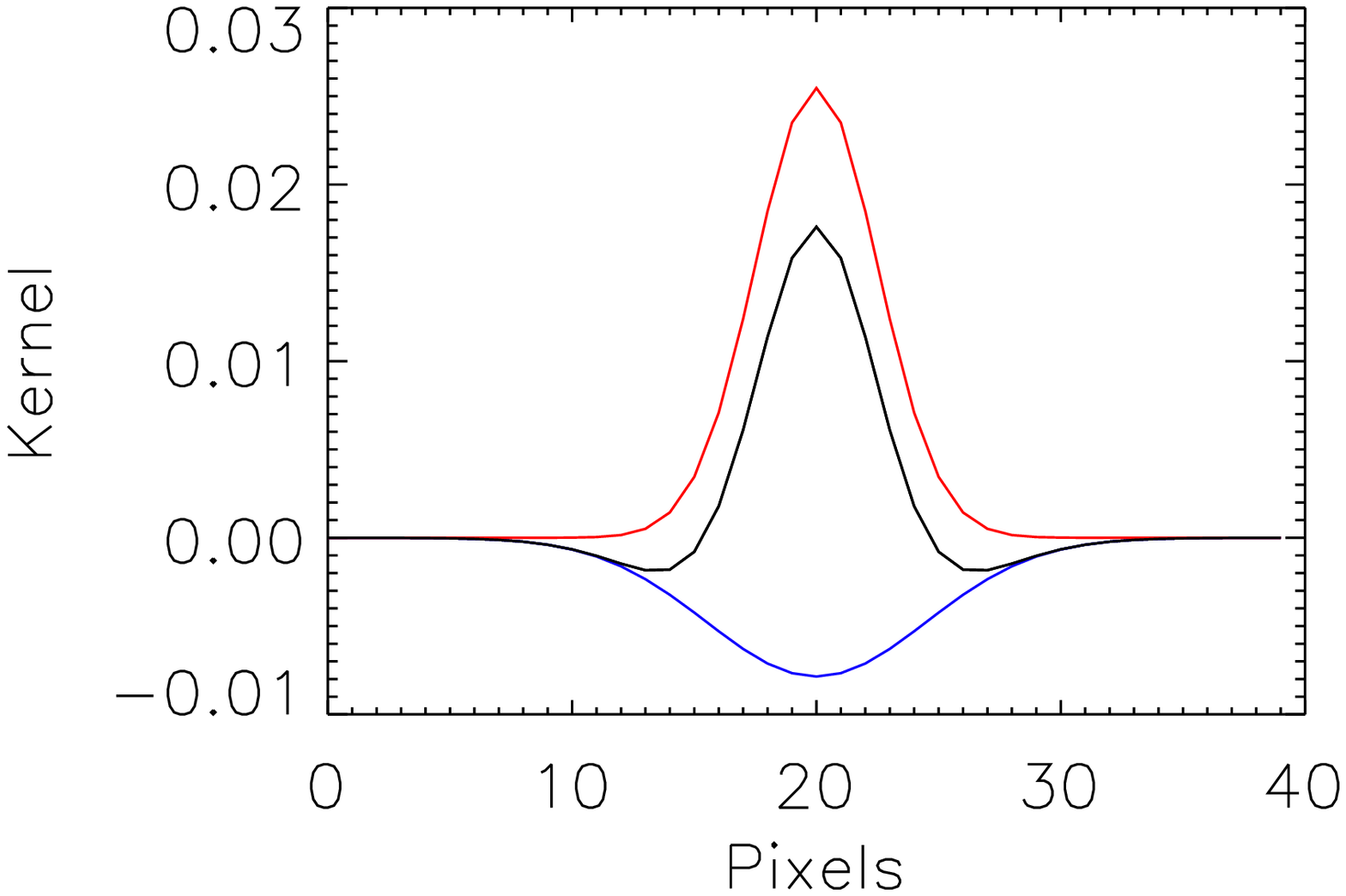}
\caption{
Profile of the difference of Gaussians kernel (black curve) that is convolved with the original image to produce a filtered image with the large scale background removed, and the visibility of the filaments enhanced. The kernel is the difference of a narrow Gaussian ($\sigma_1 = 2.5$ pixels; red curve) and a wider Gaussian ($\sigma_2 = 4.5$ pixels; blue curve).
}
\end{figure}

\begin{figure}[ht!]
\plotone{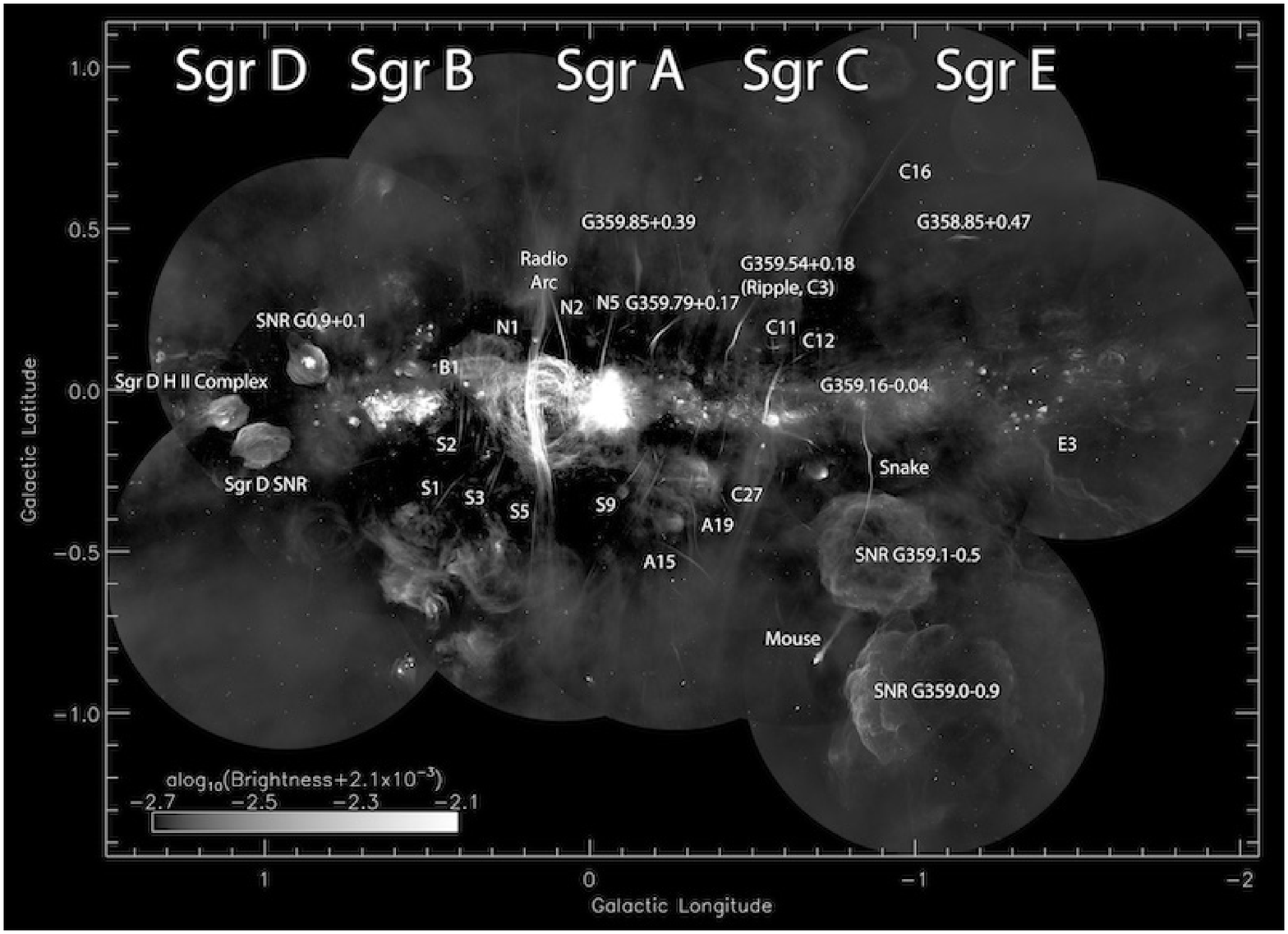}
\plotone{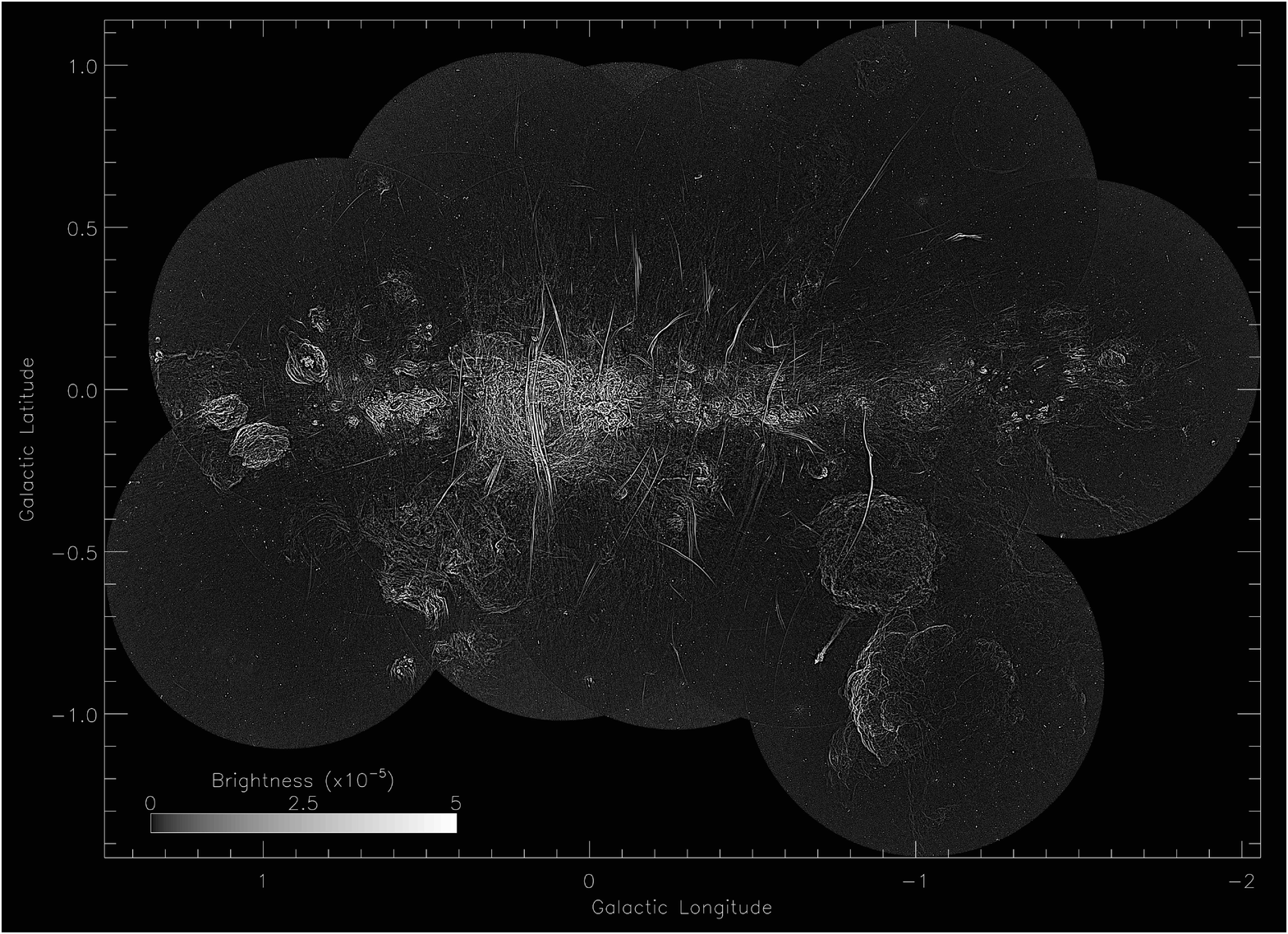}
\caption{
{\it Top 2a} 
A mosaic image of the Galactic center at 20cm with a 4$''$ resolution taken with the MeerKAT (Heywood et al. 2021). 
Prominent sources are labelled. The names of the  filaments are  taken from \citep{zadeh04,law08,larosa01}. The rms noise in this image $\sim80$ mJy beam$^{-1}$. This data product is 
publicly available in Heywood et al. (2021). 
{\it Bottom 2b} 
A filtered version of (a) with rms sensitivity of 1.8-2.3$\mu$Jy and a resolution of 6.4$''$.
The units of the  filtered image do  not account for 
the changed beam size. The subtraction built into the filtering reduces the apparent intensities by a factor of $\sim0.13$.
\label{fig:general}}
\end{figure}

\begin{figure}[ht!]
\plotone{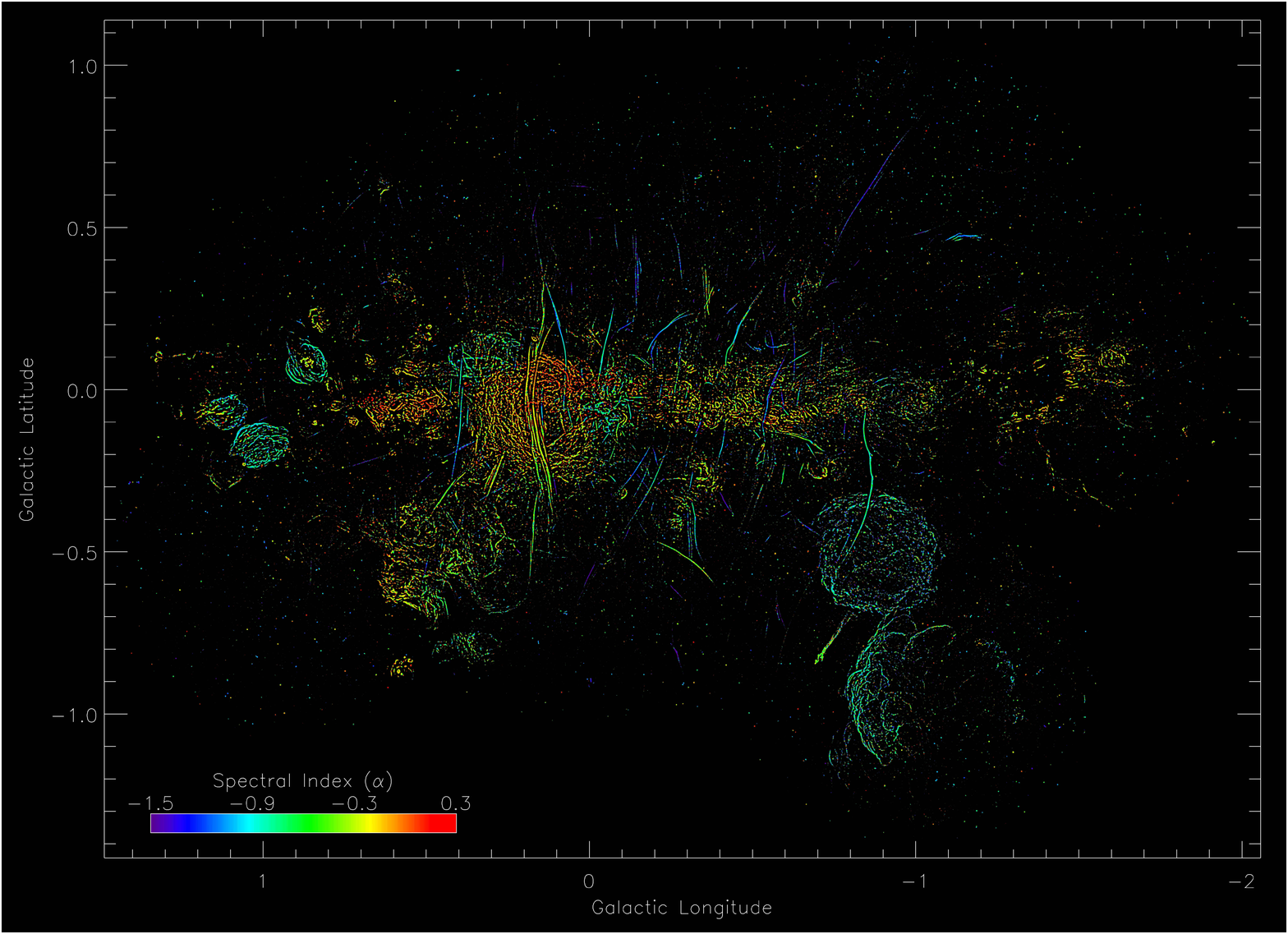}
\plotone{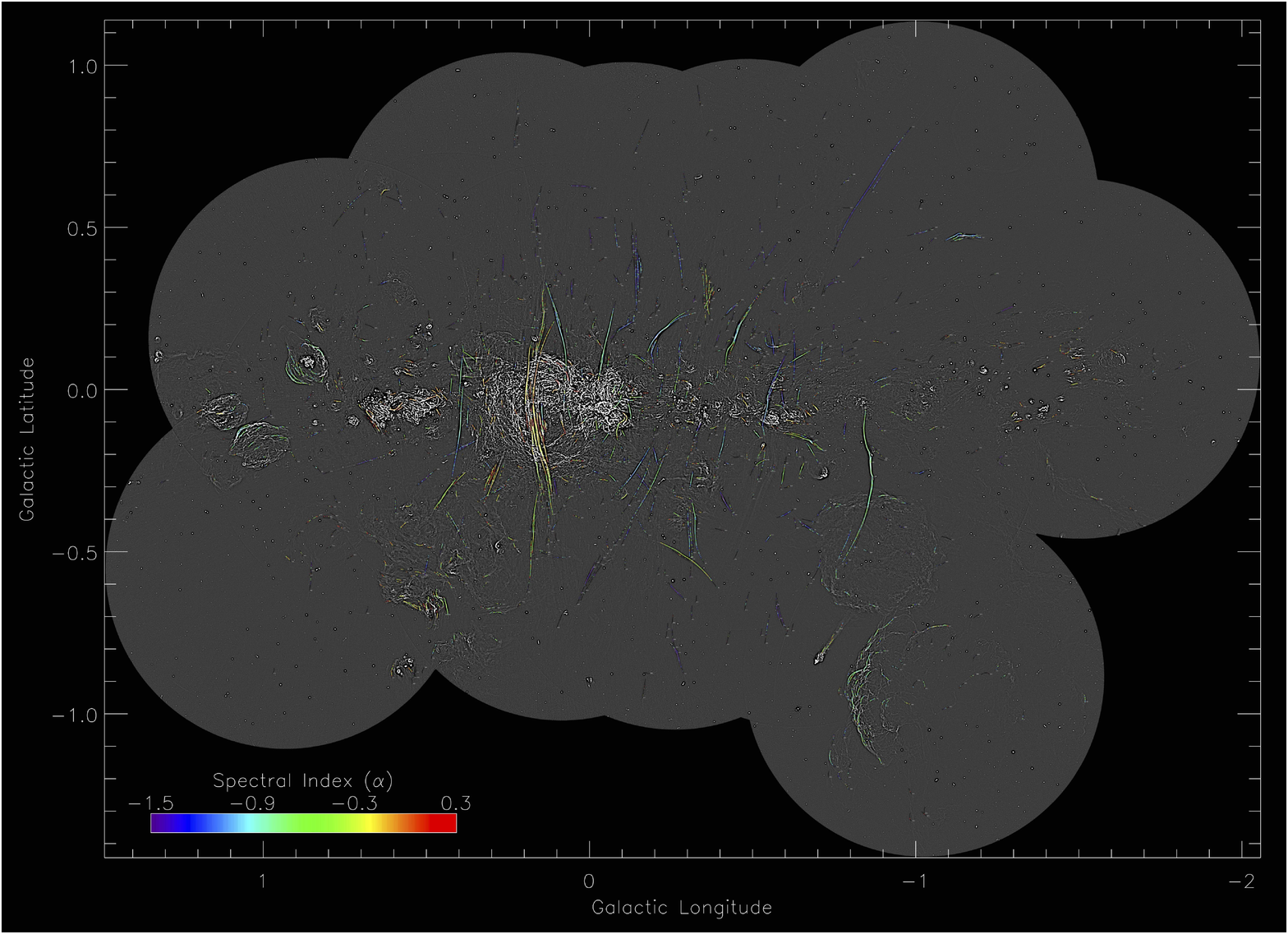}
\caption{
{\it Top  a} 
An image of the  spectral index 
for pixels with sufficient signal in the filtered image(s) to yield a reliable estimate.
{\it Bottom  b} 
An image showing the spectral index (see color bar) for filaments longer than $132''$. 
For reference, the spectral index values are superimposed on a grayscale image of the intensity.
\label{fig:general}}
\end{figure}

\begin{figure}[ht!]
\plottwo{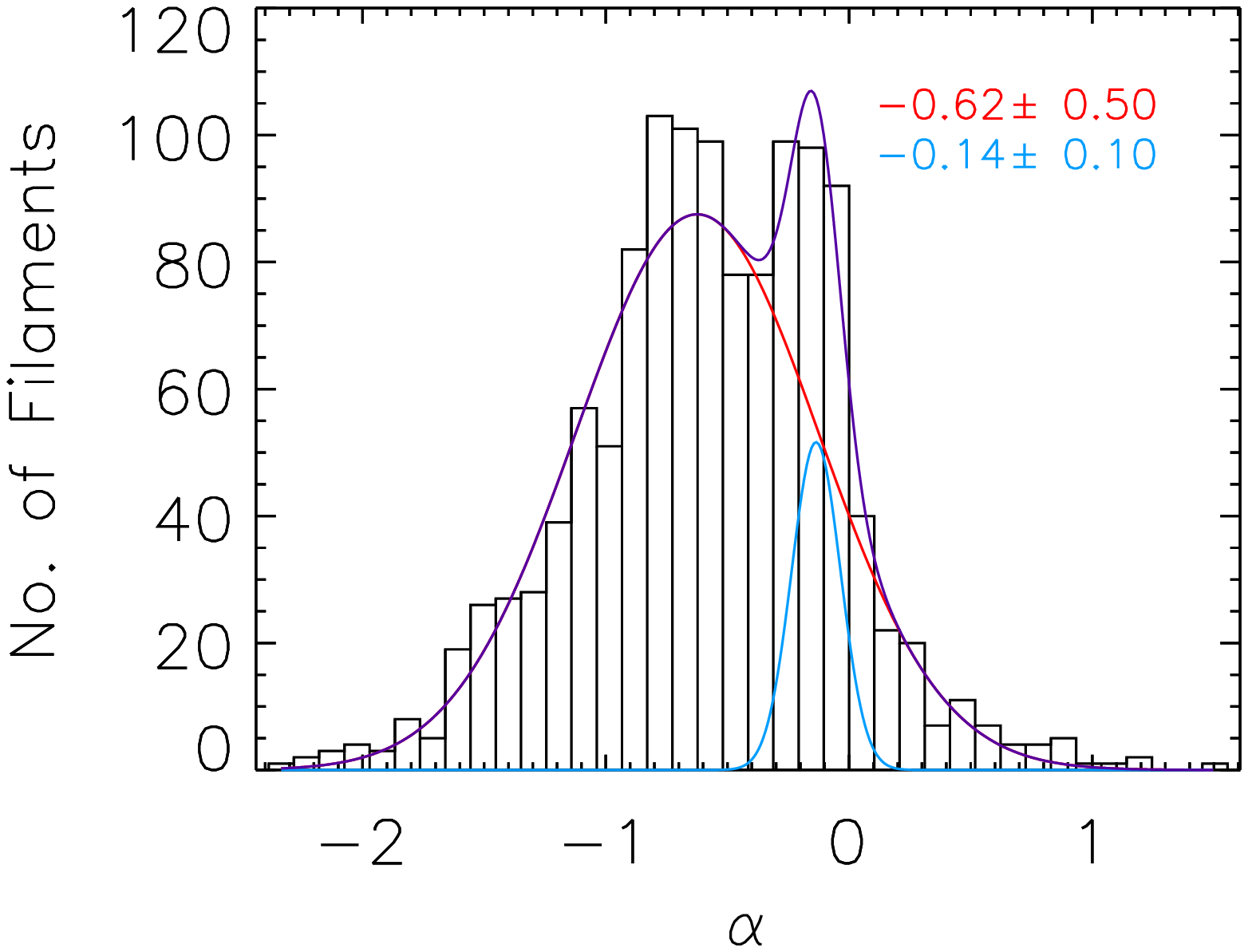}{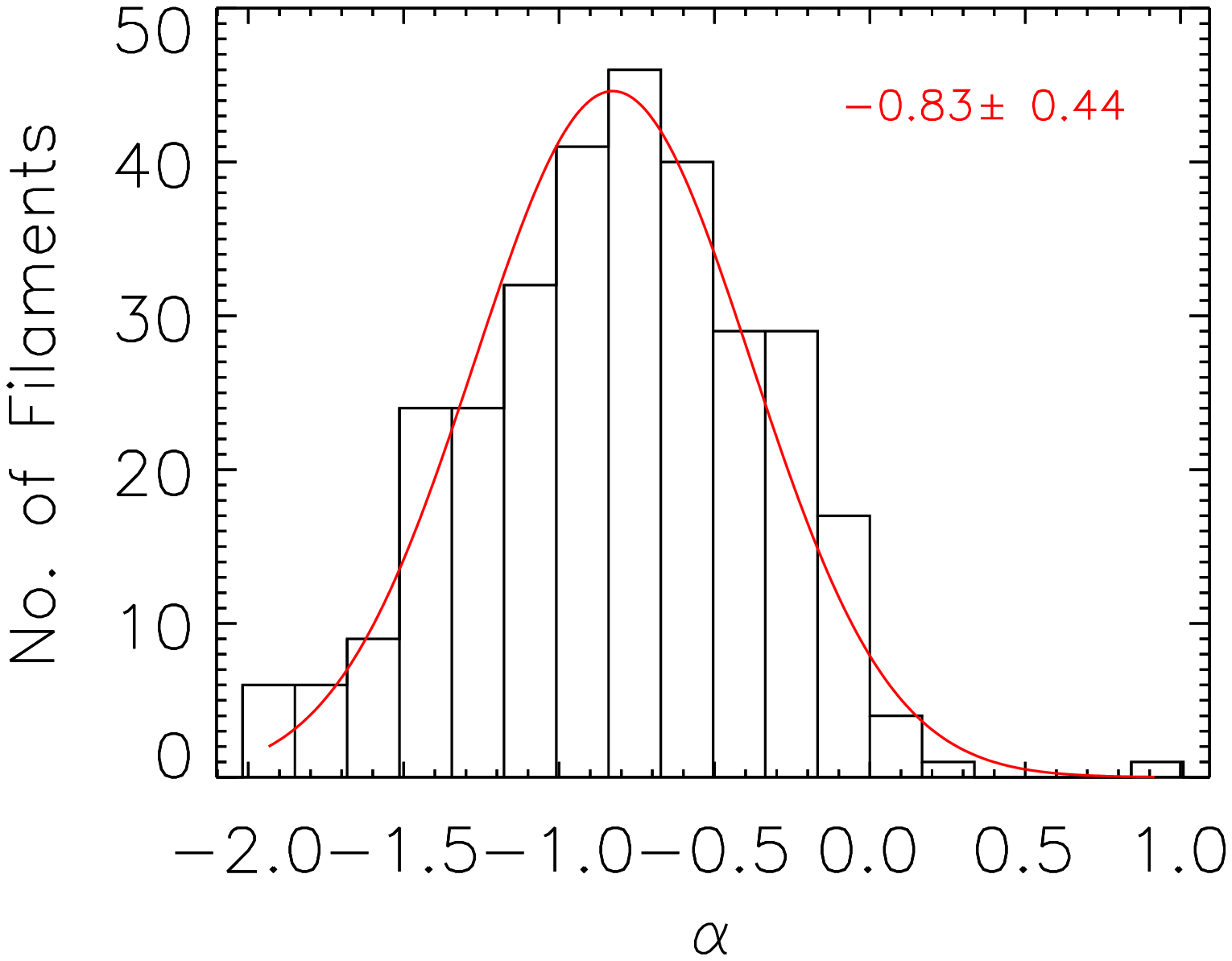}
\caption{
{\it Left a} 
Histogram of the number of filaments as a function of spectral index for linear features longer than $66''$. 
Two Gaussian components were fit to the data suggesting steep spectrum nonthermal linear 
features (red) and flat thermal HII regions (light blue). {\it Right b} Similar to (a) except long filaments 
$>132''$ were selected. \label{fig:general}} 
\end{figure}

\begin{figure}[ht!]
\plottwo{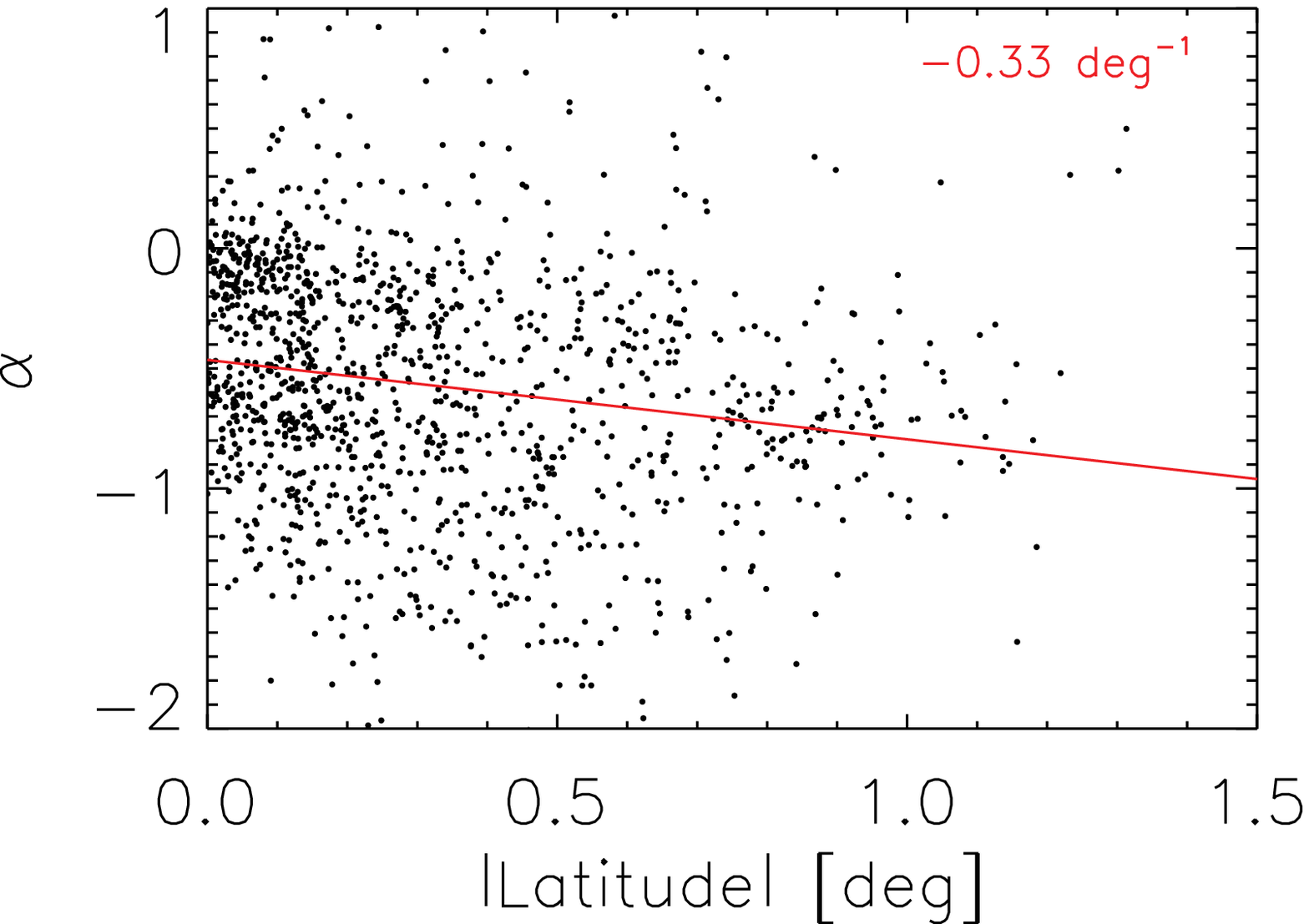}{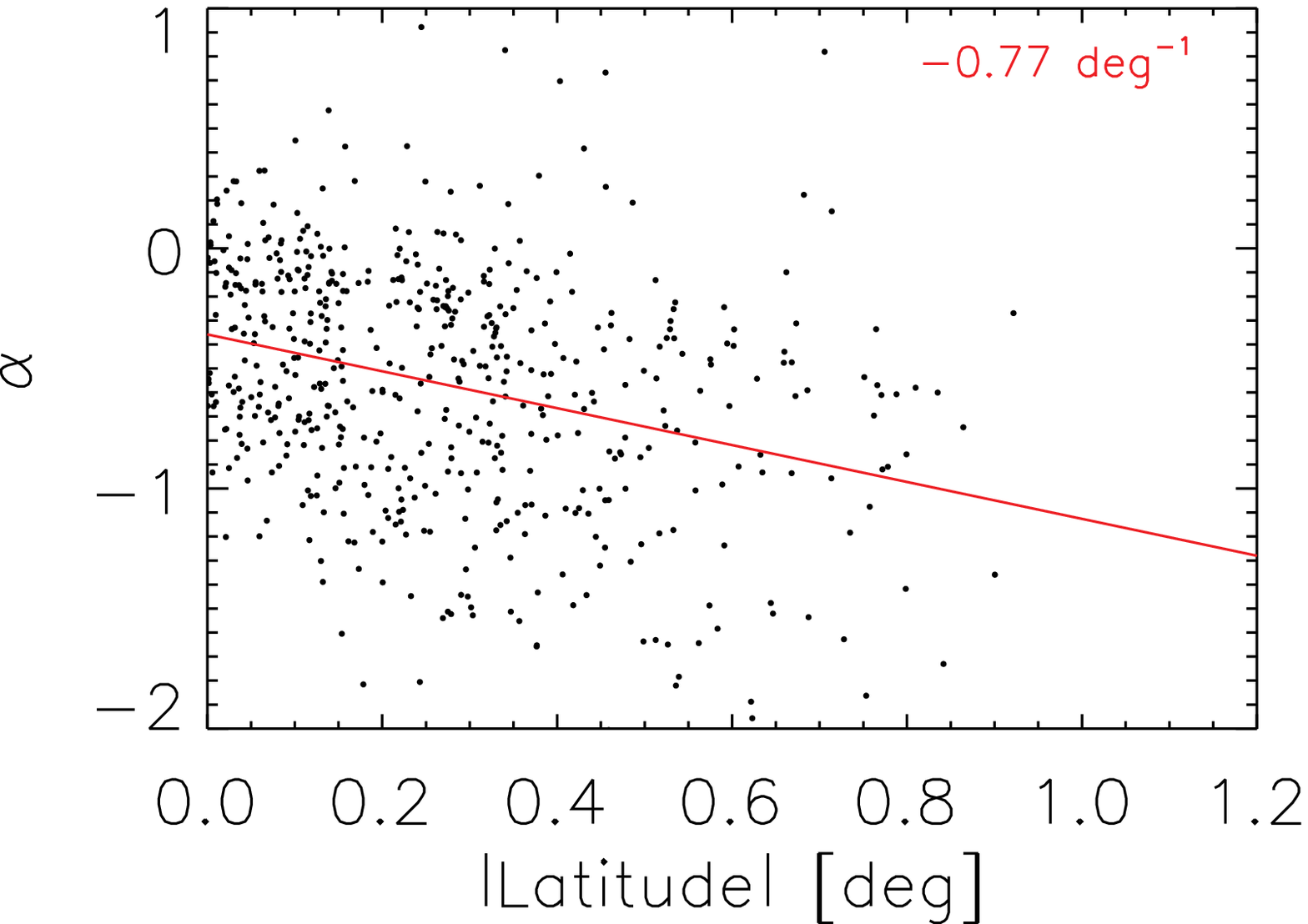}
\caption{
{\it Left a} 
A scatter plot of the spectral index as a function of absolute values of latitude 
for linear features longer than $66''$.  The red line is a $\chi^2$-fit to the data with a slope shown to the top right. 
{\it Right  b} 
Similar to (a) except only for filaments longer  than 132$''$ were selected, to exclude virtually all H II structures and most SNR shock structures. 
\label{fig:general}}
\end{figure}

\begin{figure}[ht!]
\plotone{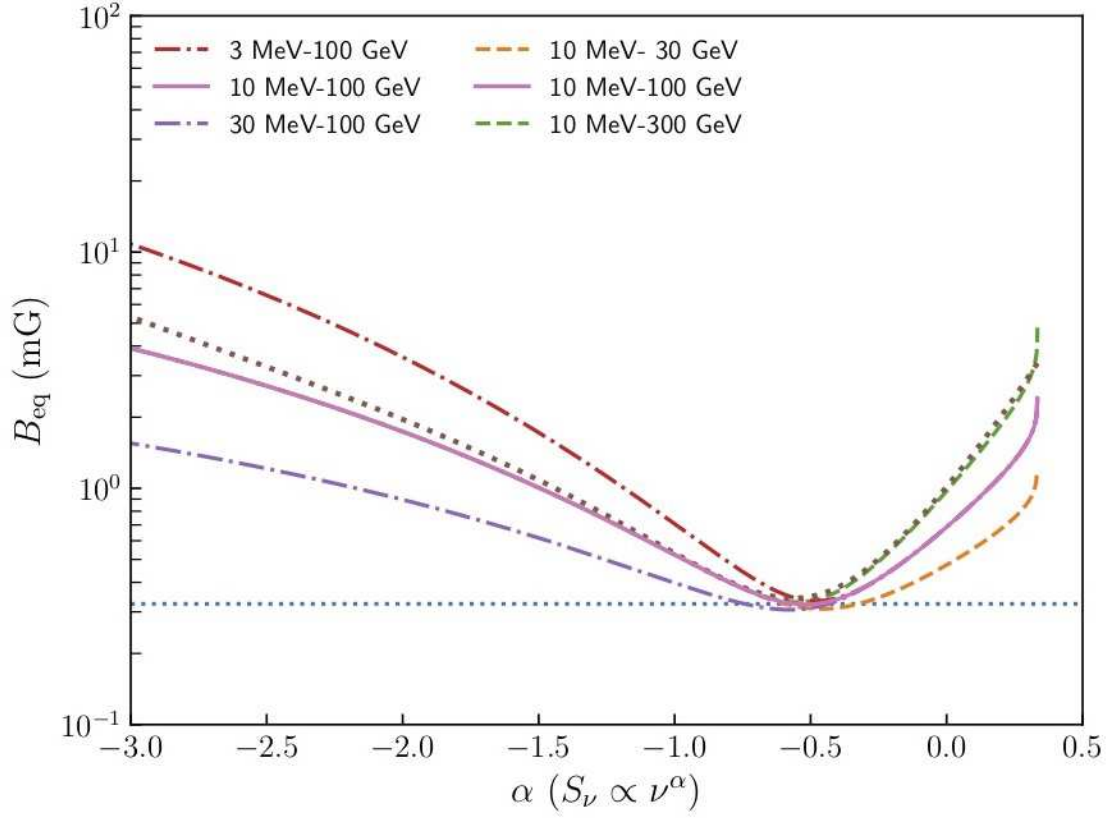}
\caption{
The  equipartition magnetic field  as a function of  spectral index using different ranges of 
electron energies.  
The dotted blue lines shows the result when    it is assumed that $\alpha = -0.5$ (i.e. Eq (1) instead of Eq. (2)). 
Plotted results corresponds to brightness $I_\nu = 1$ mJy asec$^2$ and thickness L = 1$''$.
\label{fig:general}}
\end{figure}

\begin{figure}[ht!]
\plotone{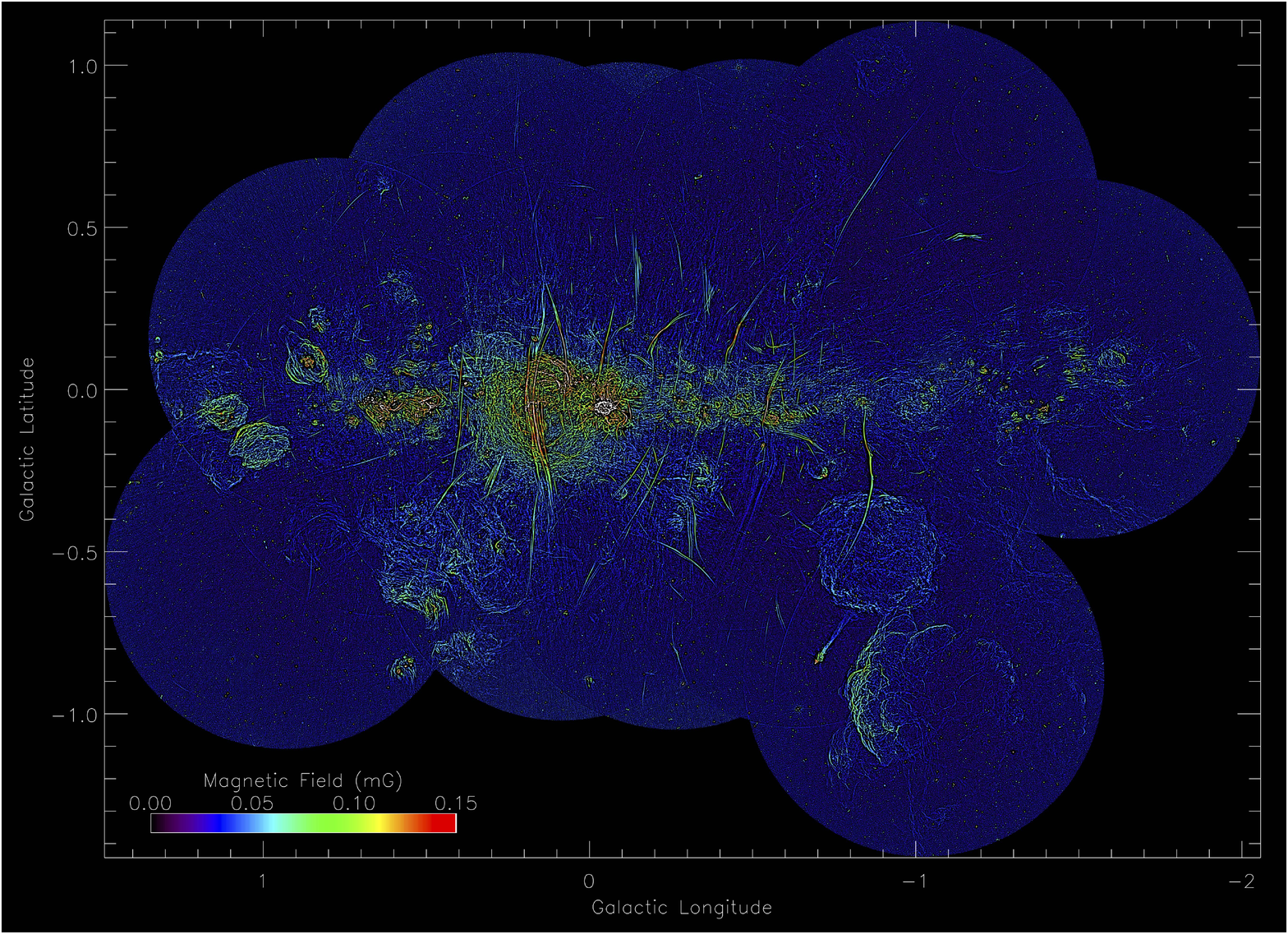}
\plotone{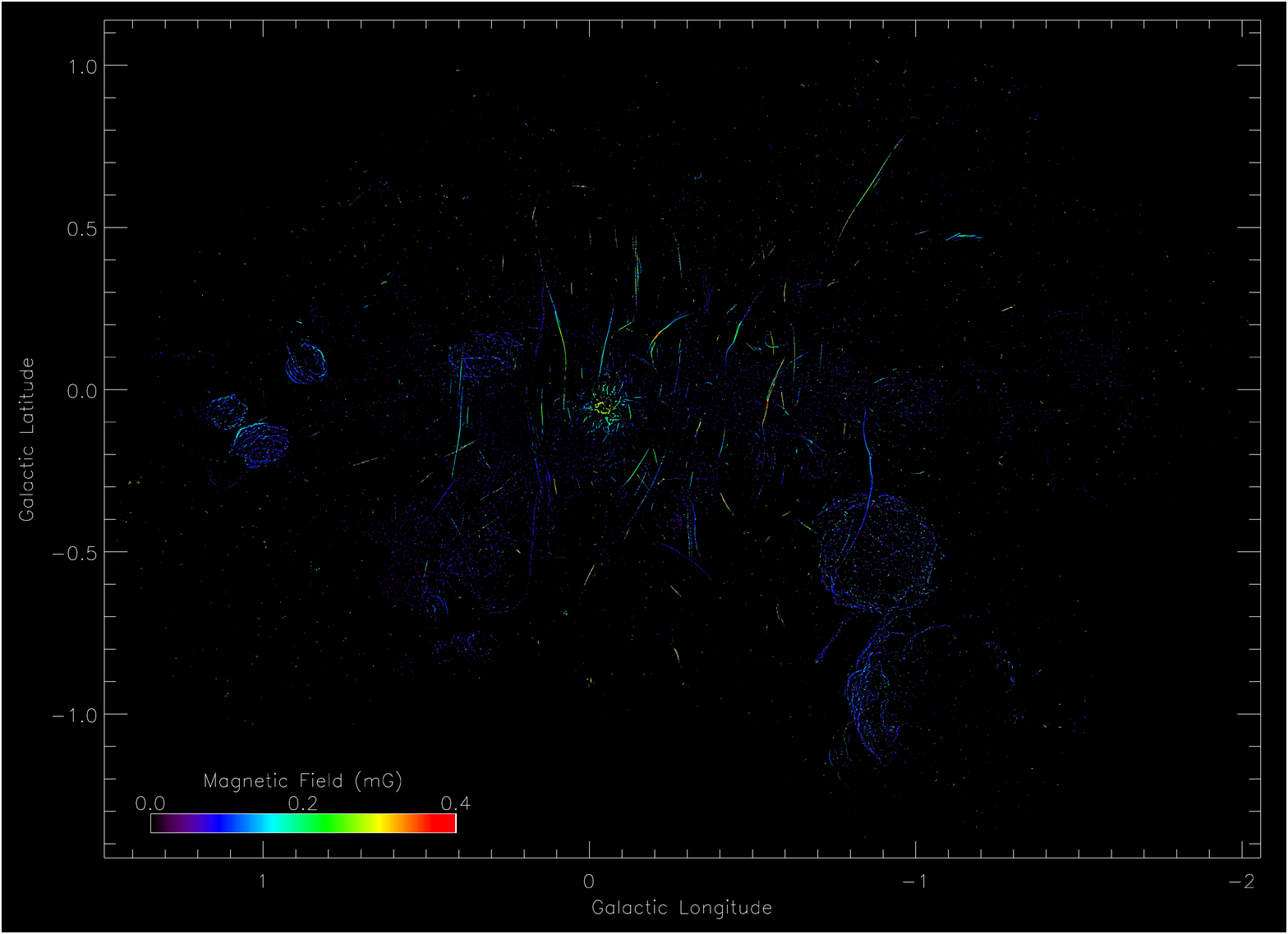}
\caption{
{\it Top  a} 
An image of the distribution of the equipartition magnetic field assuming all the features are nonthermal emission 
and have a spectral index $\alpha = -0.5$, as calculated from Eq. (1). 
{\it Bottom  b} 
Equipartition magnetic field calculated from Eq. (2), employing the measured spectral indices. In this image
features with  $\alpha>-0.4$ have been masked. 
\label{fig:general}}
\end{figure}


\begin{figure}[ht!]
\plottwo{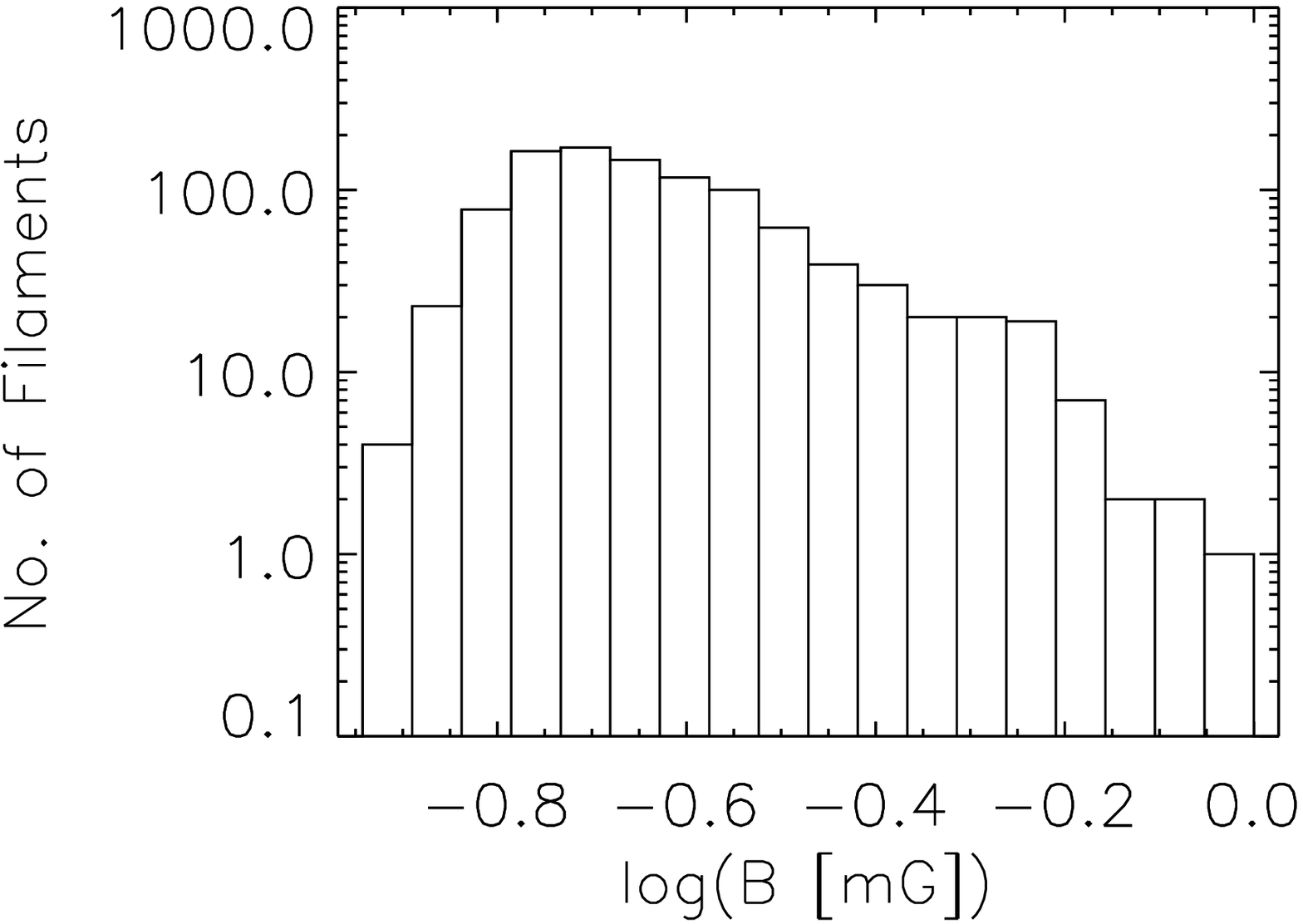}{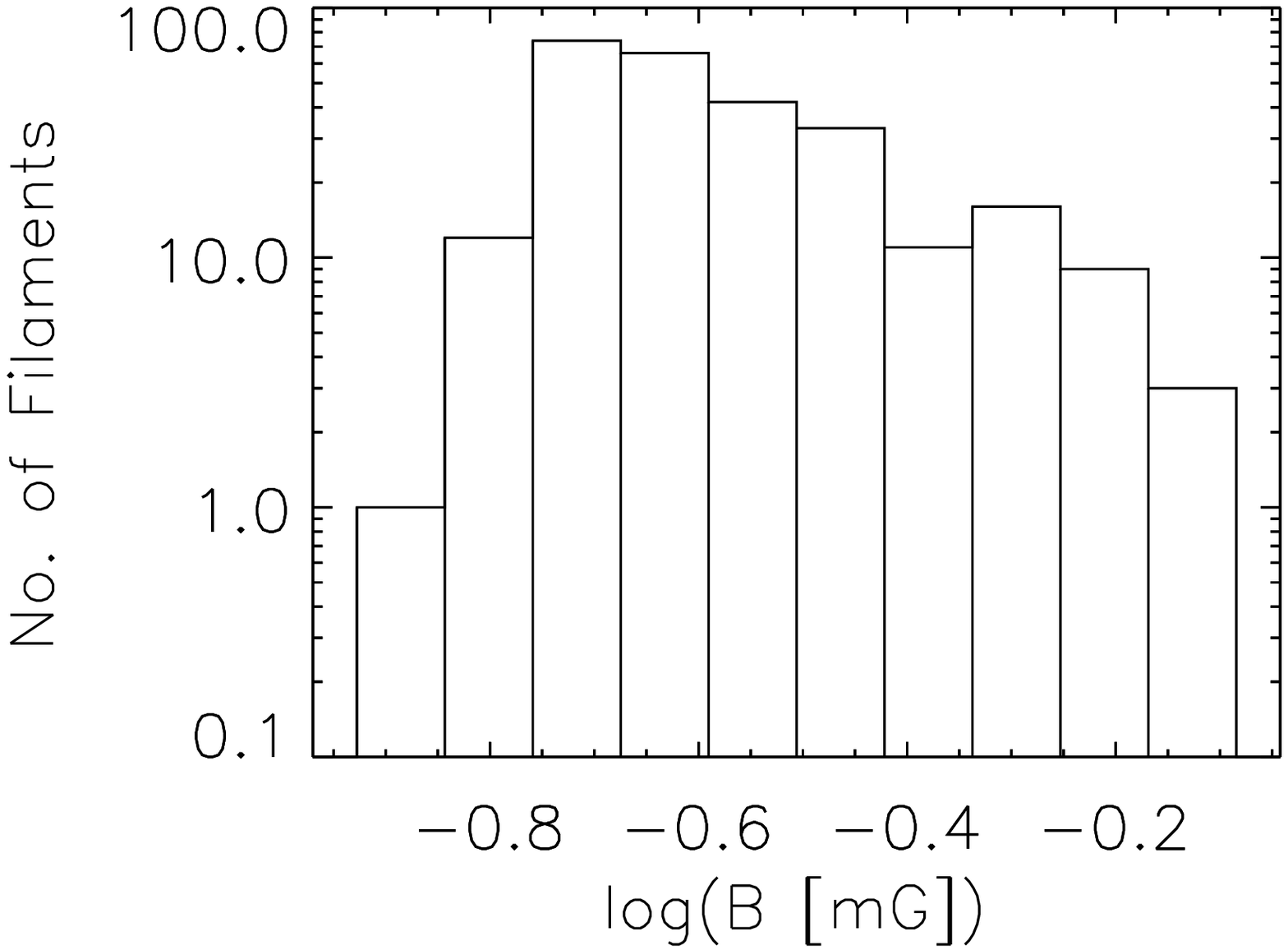}
\caption{
{\it Left a} 
Histogram of the number of filaments as a function of the magnetic field 
for linear features longer than $66''$. The variation of spectral index  has been accounted for.
{\it Right b} 
Similar to (a) except long filaments $> 132''$. 
}
\label{fig:general}
\end{figure}

\begin{figure}[ht!]
\plottwo{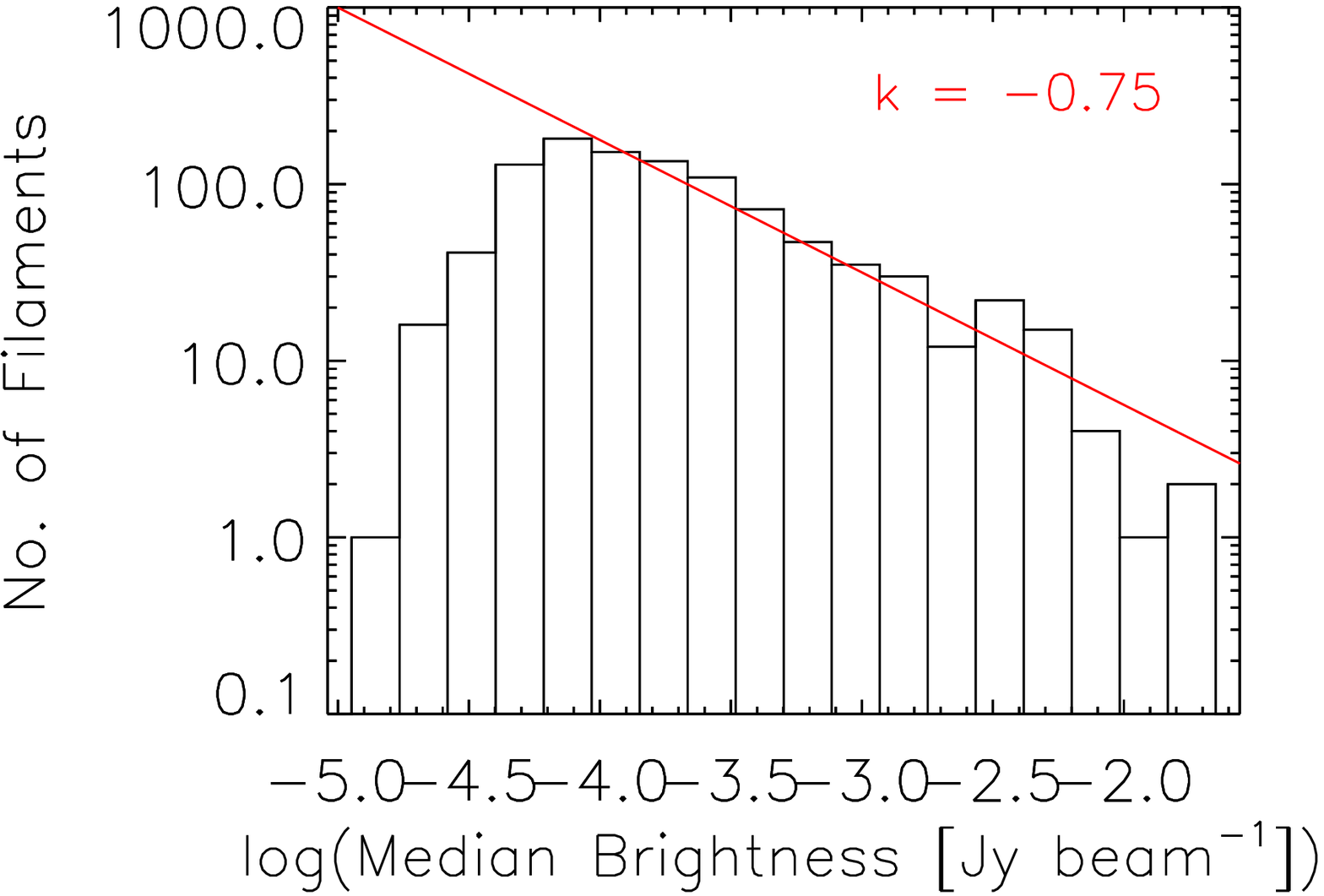}{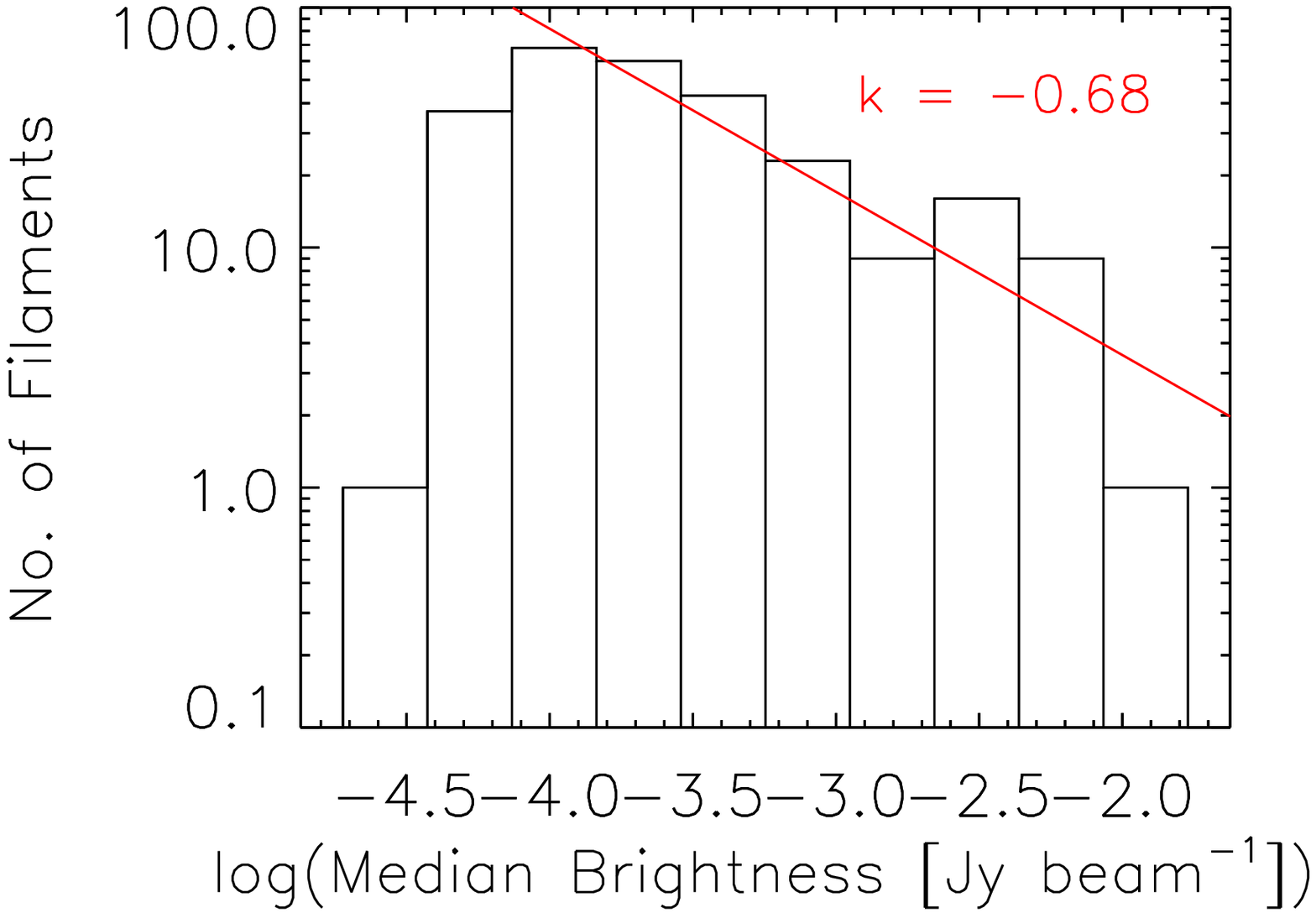}
\caption{
{\it Left a} 
Histogram of the number of filaments as a function of median surface
brightness for linear features longer than $66''$. The red line indicates a
power law fit to the bright portion of the histogram, indicating a power law
index of $k=-0.75$. 
{\it Right b} 
Similar to (a) except long filaments $> 132''$ were
selected to exclude virtually all H II structures and most SNR shock structures.
}
\label{fig:general}
\end{figure}

\end{document}